%% file: main.tex
\documentclass[pdflatex,sn-basic]{sn-jnl}

\usepackage{graphicx}%
\usepackage{multirow}%
\usepackage{amsmath,amssymb,amsfonts}%
\usepackage{amsthm}%
\usepackage{mathrsfs}%
\usepackage[title]{appendix}%
\usepackage{xcolor}%
\usepackage{textcomp}%
\usepackage{manyfoot}%
\usepackage{booktabs}%
\usepackage{algorithm}%
\usepackage{algorithmicx}%
\usepackage{algpseudocode}%
\usepackage{listings}%

\usepackage[switch]{lineno}

\usepackage{tabularx}

\newcommand{\cristysbox}[1]{

\begin{fancy}[]{customcolor} 
\noindent\parbox{0.96\linewidth}
{\vspace{1px}
#1
\vspace{5px}} 
\vspace{-5pt}
\end{fancy}
}

\usepackage{graphicx} 
\usepackage[framemethod=TikZ]{mdframed}
\usepackage{xcolor} 
\definecolor{quote}{HTML}{9673A6}

\newcommand{\pquote}[1]{\textcolor[HTML]{9673A6}{\textbf{Participant #1:} }}

\newcommand{\ta}[1]{\textcolor[HTML]{3498D4}{\textbf{#1} }}

\newenvironment{zeroindent}
  {\par\setlength{\parindent}{0pt}}
  {\par}

\definecolor{customcolor}{HTML}{79D0D9}

\newenvironment{fancy}[2][]{
    \mdfsetup{
        skipabove=1pt, 
        innerlinewidth=1pt, innerlinecolor=#2, 
        linewidth=0pt,
        backgroundcolor=#2!20 
    }
    \begin{mdframed}}
    {\end{mdframed}}

\newenvironment{coloredframe}[2][]{
    \mdfsetup{
        skipabove=2pt, 
        hidealllines=true, leftline=true,      
        innerlinewidth=2pt, innerlinecolor=#2, 
        linewidth=0pt,
        backgroundcolor=#2!10
    }
    \begin{mdframed}}
    {\end{mdframed}}

\newcommand{\dialoguegpt}[2]{
    \begin{coloredframe}{#1}
    \vspace{1px}
    \small 
    \begin{zeroindent} #2 \end{zeroindent}
    \vspace{1px}
    \end{coloredframe}
    \vspace{-5px}
}

\pdfoutput=1 

\nolinenumbers

\theoremstyle{thmstyleone}%
\theoremstyle{thmstyletwo}%

\theoremstyle{thmstylethree}%

\begin{document}
\let\cite\citep 

\title{Exploring Emotional Intelligence in Software Testing}


\author*[3]{\fnm{Cristina} \sur{Martinez Montes}}\email{cristina.martinezmontes@miun.se}

\author[1,2]{\fnm{Pushpamalar } \sur{Rajendran}}\email{rajmalar717@gmail.com}

\author[1,2]{\fnm{Grace} \sur{Karuri}}\email{gracewanjiku71@gmail.com}

\author[3]{\fnm{Felix} \sur{Dobslaw}}\email{felix.dobslaw@miun.se}

\affil*[1]{\orgdiv{Computer Science and Engineering}, \orgname{Chalmers University of Technology}, \orgaddress{\street{Teknikgatan}, \city{Gothenburg}, \postcode{41756},  \country{Sweden}}}

\affil[2]{\orgdiv{Computer Science and Engineering}, \orgname{University of Gothenburg}, \orgaddress{\street{Teknikgatan}, \city{\textbf{Gothenburg}}, \postcode{41756}, \country{Sweden}}}

\affil[3]{\orgdiv{Communication, Quality Technology and Information Systems}, \orgname{Mid Sweden University}, \orgaddress{\street{Akademigatan}, \city{Östersund}, \postcode{83140},  \country{Sweden}}}


\abstract{

\textbf{Background:} Emotional Intelligence (EI) is the ability to recognise, understand, and manage one's own and others' emotions. Software testers deliver judgements about colleagues' work under deadlines they do not control, and prior work on emotion in software engineering has mostly studied developers.

\textbf{Aims:} To explore how software testers describe the part EI plays in their day-to-day work, in communication and conflict within the team, and in responding to requirements volatility.

\textbf{Method:} Semi-structured interviews with 16 software testers in Sweden working in teams that use agile practices, across aviation, automotive, healthcare, IT services, administration, banking and pharmaceuticals, analysed with reflexive thematic analysis informed by Goleman's EI framework.

\textbf{Results:} Three themes. Testers described regulating stress under deadline pressure and drawing motivation from recognition, clarity and autonomy; managing the daily delivery of critical findings to colleagues so that trust survives; and responding to requirements change with frustration that turned into decisions about what to leave untested, into advocacy for process change, or into workarounds. Read against developer-focused studies, the themes point to features of the testing role: the work product is a criticism of a colleague's work, success is invisible while failure is attributed, and the tester's window shrinks with every upstream delay.

\textbf{Conclusions:} For testers, managing emotions is a constant job requirement. The results highlight that the importance of EI increases when the development process lacks an independent testing phase. The findings also inform implications for teams and, ultimately, for organisations and future research. }




\maketitle

\section{Introduction}\label{sec1}

The role of software testers is essential in software development to mediate between stakeholders and customers on the one side and the developers on the other side \cite{ref28}. It requires mastering multiple types of interactions, demanding skills in social, communication, cognitive, and technical areas \cite{ref37}. There is a natural tension between the role of software testers who attempt to assure --and developers who implement-- desired and often ambiguously specified or even underspecified behaviour. Miscommunication and misunderstandings risk leading to conflicts, particularly with the testers' focus on user requirements and the developers' focus on technical details \cite{ref15}.

Project challenges such as requirements volatility disrupt workflows and trigger strong emotional responses, including anger, anxiety, and fatigue \cite{ref26}. Particularly when testing complete systems end-to-end, task complexity impacts the emotional load \cite{ref7} such as keeping integrated testing information up to date, coordinating with multiple development teams and fostering team cooperation, while managing limited testing time.

Emotional Intelligence (EI), the ability to recognise, understand, and manage one's own and others' emotions, can help address these challenges and regulate behaviour \cite{ref48}.~Various studies \cite{ref3, ref91, ref92} indicate the impact of EI on teamwork effectiveness, trust-building, productivity, and contextual performance. Managing emotions is essential for enhancing work output, maintaining project momentum, and sustaining team morale \cite{ref5, ref26,ref27}.

EI overlaps with personality traits such as emotional stability \cite{ref75}, which is one reason we treat it in this study as an interpretive lens on what testers describe (Section~\ref{sec:limitations}). Within software engineering, EI has been linked to communication, performance and stress management \cite{rezvani2019emotional}. However, there is a lack of studies specifically focusing on the impact of software testers EI on software testing and in extension quality assurance. Combining the demand of incorporating and navigating developers and stakeholders perspectives, it is central to understand what skills testers particularly benefit from for their tasks. This research gap prompts our study to explore this impact with participants from Sweden, a country with comparatively high work autonomy and attention to work quality \cite{ref76}. To guide our investigation, we pose the following research questions:\\

\textbf{RQ1. How does emotional intelligence influence software testers’ experiences and practices in their work?}\\

\textbf{RQ2. How does the individual's emotional intelligence of software testers shape communication and conflict management within their teams?}\\

\textbf{RQ3. How do testers describe their emotional responses to requirements volatility, and which decisions about testing do they
connect to them?}\\

This article makes four contributions.
\begin{enumerate}
\item Three themes, generated through reflexive thematic analysis of interviews with 16 software testers in Sweden, describing how testers draw on emotional intelligence in day-to-day testing practice, in communication and conflict within the team, and in responding to requirements volatility (Section~\ref{sec:results}).
\item A reading of each theme against the developer-focused literature on emotion in software engineering, stating what the testing role adds (Section~\ref{sec:tester-specificity}). The one with the most direct bearing on software quality is that a tester's emotional response to requirements change converts into a decision about what is left untested.
\item A proposition on why emotional and interpersonal competence carries more weight for testers once development has no separate test phase (Section~\ref{sec:agile-conjecture}), and a discussion of how EI scales from individual coping to a team and organisational capability.
\item Implications for practice aimed at testers, team leads and organisations.
\end{enumerate}

Our study aims to understand the impact of emotional aspects on the workdays of software testers. Specifically, it focuses on how EI shapes software testers approach to ensuring software quality through self-awareness, self-management, social awareness, and relationship management \cite{ref4}. 

To achieve the objectives of this study, we conducted interviews with 16 software testers of varying experience levels who worked in agile settings across different industries, including aviation, automotive, healthcare, IT services, administration, banking, and pharmaceuticals in Sweden. The participants were engaged in either manual testing or test automation.

\input{Background}

\input{Related_work}

\input{Methods}

\input{Results}

\input{Discussions}

\input{Conclusion}

\section{Statements and Declarations}

\textbf{Author Contributions:}
CMM, PR and GK developed the main idea and concept of the paper. CMM, PR, GK and FD performed formal analysis and investigation. CMM and FD wrote, reviewed and edited the manuscript.

\textbf{Funding:} The authors did not receive support from any organisation for the submitted work.

\textbf{Data Availability:}
The dataset Exploring Emotional Intelligence in Software Testing, which includes the codes and themes supporting this study's findings, is publicly available on Zenodo at https://doi.org/10.5281/zenodo.20271239. The interview transcripts are not publicly available because doing so could compromise the anonymity and confidentiality of the participants.

\textbf{Competing Interests:} The authors have no relevant financial or non-financial interests to disclose.

\bibliography{references}

\end{document}

%% file: Background.tex
\section{Background}
In this section, we elaborate on the main concepts and approaches considered in the study.
\subsection{Software Testing in Contemporary Development}
\label{sec:agile-conjecture}
Software testing ensures software quality by verifying that software products meet specified requirements, thus contributing to reliable and robust systems \cite{ref2,ref36}. In contemporary development, testing is integrated throughout the lifecycle, which requires close collaboration between testers, developers, and stakeholders from the earliest stages, including the creation of requirements \cite{ref73,ref78}. The continuous involvement aims to facilitate the co-development of relevant, non-redundant test suites and to improve defect detection efficiency \cite{ref16,ref56,ref66}. Many real-world software systems consist of substantial lines of testing code, often as much as production code.

Most software today is developed iteratively, in cross-functional teams that deliver in short cycles and share responsibility for quality. Testing is integrated throughout development rather than saved for a separate final phase. We use the term agile practices for this way of working, whatever the named process behind it (Scrum, Kanban, scaled frameworks, or a hybrid with a phased model). Because our participants described their work this way, our study focuses on testing under these conditions, not on any particular process.

Despite its advantages, testing under agile practices presents organisational and interpersonal challenges. Testers can face time pressure, limited support, and difficulty establishing their role in teams accustomed to traditional models \cite{deak2016challenges,ref99}, while often being held responsible for undetected software failures and therefore placed in a vulnerable position. Effective collaboration, communication, and emotional intelligence are thus essential to sustaining high-quality outcomes in agile settings \cite{ref23}.

The disappearance of the separate test phase also changes where the testing role draws its authority from. Under a phased process, the test phase acted as a formal gate, and the tester's standing to hold or delay a release was written into the process. Iterative delivery distributes quality across the whole team, developers write and own tests, and that gate is gone. We therefore proceed from the following conjecture, which we state as a premise: once the structural authority of the testing role is removed, a tester's influence over quality decisions becomes largely relational, so emotional and interpersonal competence carries more weight for testers today than it did under a phased process. Section~\ref{sec:tester-specificity} returns to this conjecture in the light of what our participants described.

\subsection{Emotional Intelligence}
Emotional Intelligence (EI) is the ability to recognise, understand, and manage one's own and others' emotions \cite{ref48}.~In 1990, Peter Salovey and John Mayer first proposed their theory of EI \cite{ref13}.~Over the following decade, theorists developed several distinctive EI models \cite{ref13, ref20, ref4}, all building on this foundational definition of EI.

There are three well-known EI models in the literature: the Ability Model, the Bar-On emotional-social intelligence model, and the Goleman performance-based competency model~\cite{bar2006bar}. The Ability Model, originally proposed by \citet{ref13}, conceptualises EI as a set of cognitive abilities related to perceiving, using, understanding, and managing emotions. The model by \citet{ref4} expanded this conception and incorporated personality traits, behavioural competencies, and emotional abilities that drive managerial performance. The Bar-On emotional-social intelligence model~\cite{bar2006bar}, refers to a set of interconnected emotional and social competencies, skills, and enabling factors that influence intelligent behaviour. While these models differ in their theoretical foundations, they share a common emphasis on the role emotions have in shaping thought and behaviour.

EI has been linked to positive outcomes in collaborative and organisational settings. For example, studies such as the one by \citet{jordan2021managing} found that EI indicators are positively associated with team performance and conflict resolution methods. Beyond individual and team impact, \citet{coronado2023emotional} found that leaders with high EI had a positive impact on behaviours and business results, as well as on team performance, including team members’ attitudes about work. These characteristics make EI particularly important in software engineering roles, especially agile software development and testing.

Agile methodologies place significant emphasis on communication, adaptability, and close collaboration within cross-functional teams. Software testing in agile environments further demands that practitioners manage uncertainty, respond constructively to feedback, and maintain productive working relationships under pressure. Given these work settings, EI competencies are likely to influence individual performance and team effectiveness in agile testing contexts.

\subsubsection{Goleman's Emotional Intelligence Framework.}
We chose the Goleman framework \cite{ref4} over the rest because it is an EI-based theory of performance that directly applies to the domain of work and organisational effectiveness.~According to this framework, an individual's potential to master skills across four domains can significantly impact success in the workplace.~

These domains are: Self-Awareness, Self-Management, Social Awareness, and Relationship Management.

Each domain contains several EI competencies. Self-awareness and Self-management are centred on the self, whereas social awareness and Relationship management focus on interactions with others. Recognition involves identifying and comprehending emotions in self-awareness and social awareness domains. Regulation pertains to the management of emotions and is closely linked to self-management and relationship management. Table \ref{tab:goleman_domains} shows an overview of Goleman's domains and competencies.

Software testing in general and agile software testing in particular are a collaborative, continuous practice shaped by interpersonal challenges, where EI influences how effectively individuals and teams operate. Hence, it is important to assess how this influence has been observed and understood in prior software engineering research. The next section explores this.

\begin{table}[ht]
\centering
\caption{Goleman's Emotional Intelligence Domains}
\label{tab:goleman_domains}
\begin{tabular}{lll}
\hline
 & \textbf{Self (Personal Competence)} & \textbf{Other (Social Competence)} \\
\hline

\textbf{Recognition} & \textbf{Self-Awareness} & \textbf{Social Awareness} \\
 & Emotional self-awareness & Empathy \\
 & Accurate self-assessment & Service orientation \\
 & Self-confidence & Organizational awareness \\

\\

\textbf{Regulation} & \textbf{Self-Management} & \textbf{Relationship Management} \\
 & Emotional self-control & Developing others \\
 & Trustworthiness & Influence \\
 & Conscientiousness & Communication \\
 & Adaptability & Conflict management \\
 & Achievement drive & Visionary leadership \\
 & Initiative & Catalyzing change \\
 &  & Building bonds \\
 &  & Teamwork and collaboration \\

\hline
\end{tabular}
\end{table}

%% file: Related_work.tex
\section{Related Work}

This section elaborates on the previous work on which we built our research.

\subsection{Emotions and EI in Software Engineering}

Several studies have explored the influence of emotions on different software engineering tasks. For example, \citet{ref3} found that software developers' productivity correlates with emotional states, with medium arousal and positive emotions linked to higher output. Aligned with this, \citet{ref5} explored the concept of emotional risk; they concluded that positive emotions, such as pleasure, enhance efficiency, while negative emotions, like frustration, can hinder performance. In another study, \citet{madampe2024supporting} showed that empathising with team members and monitoring commitments and decisions are effective EI-based strategies for managing emotions, sustaining productivity, and achieving team goals. Finally, \citet{graziotin2018happens} researched how happiness, or the lack of it, influences software development, adding to the big picture of how emotions affect each task and outcome and how essential it is for software professionals to navigate optimally the emotional complexities of their work to improve personal team well-being and performance. These studies have had software engineers as the population, without much distinction in how emotional demands may differ between roles. In the case of software testers, their work demands technical evaluations and interpersonal interactions. They need to act as a link between business and development while managing conflict when communicating defects/ bad news and disagreements \cite{ref19}. In agile settings, testers collaborate continuously with stakeholders, developers and product owners, demanding different levels of communication too. For the previous reasons, in this study, we focused on testers. 

Emotional intelligence facilitates \textbf{team collaboration and unity}, becoming essential to project performance \cite{rezvani2019emotional}. \citet{ref91} reported that employees in self-managed teams with high EI levels demonstrated greater teamwork effectiveness, making the connection between EI competencies and collaborative performance relevant. In the same way, \citet{ref92} observed a positive correlation between EI, overall performance, and organisational commitment, showing that emotionally intelligent professionals are more engaged and reliable in their work. Comparably, \citet{rezvani2019emotional} concluded that EI is used as a stress mitigator and helps to foster trust among software developers working on the same project. 

These studies demonstrate how EI facilitates empathy, conflict resolution, and team communication to achieve smoother collaboration between software professionals. Regarding an agile environment that demands continuous interaction with developers, testers, and stakeholders, EI is a main factor in maintaining cohesion, negotiating differing perspectives, and sustaining productivity across iterative cycles.

\subsection{Human Factors and EI in Software Testing}

Regarding the specifics of software testers, researchers have identified unique factors that influence their motivation, stress levels, and overall effectiveness. These factors can be divided into three main groups: organisational, interpersonal and intrapersonal \cite{guveyi2020human}. The organisational factors included time pressure, repetitive tasks, lack of recognition, and limited control over work processes \cite{deak2016challenges, ref99}. However, \citet{guveyi2020human} discovered that interpersonal and intrapersonal factors (in-team communication and the ability to work together) have a bigger influence than the organisational ones. \citet{deak2016challenges} closely looked at these aspects and focused on motivation; they reported that the lack of influence and recognition, and being unhappy with management, were the main demotivators of testers. On the contrary, the enjoyment of challenges, commitment to improving quality, and valuing recognition boosted testers' motivation. Most motivators and demotivators are related to emotions; testers must regulate stress, maintain motivation, and manage interactions constructively while coping with the frustration of being undervalued \cite{deak2016challenges}. EI can help testers to better adapt to agile practices, sustain performance under pressure, and contribute effectively to team goals. 

Despite EI being an essential tool for testers to achieve their goals and keep their well-being, the topic has not been explored in depth. This study aims to contribute by presenting the perceptions and experiences of testers about EI. With this, our goal is to present evidence on how EI contributes to mitigating the impact of everyday stressors and enables testers to leverage the rewarding aspects of their work, ultimately contributing to better individual well-being, team collaboration, and software quality.

%% file: Methods.tex
\section{Methodology}
We conducted a qualitative interview study to explore the importance and use of emotional intelligence among software testers working in teams that use agile practices, in the sense given in Section~\ref{sec:agile-conjecture}. Given our study's exploratory and interpretive nature, interviews were chosen as they allow for a rich understanding of complex human interactions and emotions. This method enabled participants to describe situations from their own perspectives and in their own words, providing insights that would be difficult to capture through other approaches \cite{ref85}.
We designed semi-structured interviews with open-ended questions to allow flexibility for follow-up questions and deeper topic exploration, and at the same time to ensure consistency across participants. To further enhance the validity and quality of our qualitative study, we adopted a sampling strategy aligned with established criteria for rigorous research. Our approach adhered to benchmarks of sensitivity to context, methodological rigour, transparency, coherence, and relevance, thereby supporting the credibility and impact of our findings \cite{ref60}.

\subsection{Target Population}
\label{sec:methods-population}
We targeted software testers working in teams that use agile practices, fully or in a hybrid with a phased model. We recruited participants purposively, paying attention to age, professional experience, and type of company to include a variety of perspectives relevant to our research questions \cite{robinson2014sampling}. We based this on the theoretical premise that software testers with more professional experience tend to encounter slightly fewer challenges over time \cite{deak2016challenges}. We sent personal invitations to candidates in software tester positions. The invitations were mainly by email. We aimed to include an even balance among the characteristics previously mentioned. However, we prioritised having variety in years of experience.

\subsection{Interview as Data Collection Tool}
We used a semi-structured interview, following a guide aligned with the research questions, the Goleman framework, and the specific emotions under investigation. We then conducted a pilot interview (not included in the final data) to ensure the integrity and clarity of the data collection process, which helped us identify one ambiguous question, leading to its revision. The final interview guide had two parts: Part 1 collected background information from the participants. Part 2 focused on the practical application of EI in the participants' environment. Table \ref{tab:interview_guide} shows the interview questions. Specific scenarios were explored, such as high-pressure deadlines, changing requirements, and handling positive and negative emotions in the workplace. The interviews were done in person, in English, ranged from 22 to 56 minutes and were audio-recorded and later transcribed for analysis.

\begin{table}[h]
\centering
\caption{Interview Guide}
\label{tab:interview_guide}
\begin{tabular}{p{0.90\linewidth}}
\hline

\textbf{Part 1: Background} \\ \hline
1. How old are you? \\
2. How many years of experience as a test engineer in Sweden do you have? \\
3. How do you think emotions and understanding people's emotions apply to software requirements? \\

\hline
\textbf{Part 2: Topic questions} \\ \hline
4. Can you recall a specific emotional experience in your testing career where you believe your emotional intelligence played a crucial role? \\
\quad -- How did this experience impact your ability to manage and navigate those emotions? \\

5. Can you recall an instance where your approach to testing was influenced by your understanding of a teammate's emotional state? \\
\quad -- How did this awareness shape the testing process? \\

6. Can you describe a time during testing when changing project requirements influenced your emotional state, particularly feelings of frustration or anger, and how this affected your decision-making regarding what to test first? \\
\quad -- Did this lead to fully meeting the revised requirements or did it primarily involve addressing edge cases? \\

7. In your day-to-day role as a test engineer, how has your approach to communicating with others in the team while seeking assistance influenced the effectiveness of your work? \\

8. Can you describe a situation in your role as a test engineer where you encountered disappointment, whether it was due to insufficient requirements, lack of support from managers or stakeholders, or any other related factors? \\
\quad -- How did this disappointment affect your approach to subsequent tasks? \\

9. On the flip side, could you share an experience where you felt pleased with your work as a test engineer, perhaps after completing a complex task or meeting stakeholder expectations? \\
\quad -- How did this sense of emotion influence your overall approach to quality assurance? \\

\hline
10. Is there anything else you'd like to add or discuss that we haven't covered, particularly regarding emotional intelligence in your role as a test engineer? \\

\hline
\end{tabular}
\end{table}

\subsection{Data Analysis}
\label{sec:methods-analysis}
Having 16 different interviewees gave us a specific angle and enough data for an exploratory study to understand, at a high level, the experiences of testers. However, despite the descriptive data we obtained, we acknowledge that the analysis and understanding developed in this study are only partial. Hence, our aim is to offer a contextualised analysis of these experiences rather than a universal claim \cite{tracy2010qualitative}.

After transcribing the interviews verbatim, we analysed them using Braun and Clarke's reflexive thematic analysis \cite{braun2021thematic}. The analysis was framework-guided: coding stayed close to participants' own words, and theme development was informed by the research questions and by Goleman's domains, as described in Section~\ref{sec:rigour}. The six steps were done iteratively as follows:

\textbf{Phase 1: Data familiarisation}

Two authors conducted every step of the data collection. From developing the interview guide, conducting the interviews, and verbatim transcribing the recordings, immersion and data familiarisation began early in the study. Later, after transcription, a third author joined in the process of reading the transcripts several times to fully understand the data. The three authors met to discuss the data and share insights from the data collection phase. The authors shared reflections, particularities from the interviews, questions and parts where the data needed clarity.

\textbf{Phase 2: Generating initial codes}

Coding was initially done by two authors simultaneously, paying equal attention to each response. Each meaningful part of the interviews was coded and later discussed. The majority of the codes were done semantically, sticking closely to testers' experiences and their own words. However, as coding progressed, all codes were revised, and some latent codes emerged. The fourth author joined to act as ``critical companion''. This helped reduce tunnel vision and raised questions to encourage reflexivity, find and challenge assumptions in the analysis and explore alternative interpretations. In this stage, the three authors discussed the data and codes and shared their emotional responses triggered by the interviewees' experiences. During coding, the authors began noticing patterns in the data; hence, memos were written at this stage as preparation for theme generation.

\textbf{Phase 3: Generating initial themes}

To generate the themes, we gathered codes and organised them in a way that reflects the stories participants expressed. We used the memos and interview transcripts to make sense and structure the story. Even though each participant had a different story, there was commonality in their experiences. All related codes were grouped into potential themes and sub-themes. We discussed each candidate theme and contrasted them against the raw data during the process. We were open, attentive, and reflective, allowing initial ideas to go if necessary.

\textbf{Phase 4: Developing and reviewing themes}
 
After generating the initial themes and sub-themes, we went back to the raw data and codes to refine them. After reviewing the data and especially the patterns we found, we proposed the final themes.

\textbf{Phase 5: Refining, defining and naming themes}

In this part of the process, we acknowledged that there were several ways to organise the themes to tell our participants' story. Therefore, in this step, we had more meetings to articulate ideas and gain more clarity. After defining the final themes, we named them and chose the quotes that best illustrate their main ideas.
 
\textbf{Phase 6: Writing the report}

Writing the report as the final step required us to revisit the data and previous phases, including the interviews, to refine and clarify the story. The discussion among the authors helped question the interpretation of the data and reflect on participants' experiences.

Table \ref{tab:coding_example} shows an example of a code and its theme.

\begin{table}[ht]
\centering
\caption{Example of coding process}
\label{tab:coding_example}
\begin{tabular}{p{5cm} p{3.8cm} p{3cm}}
\hline
\textbf{Data Segment} & \textbf{Code} & \textbf{Generated Theme} \\
\hline
They had not informed us that testers were not needed despite having asked several times. Somehow, they did not inform us that we were not within the scope of their testing. 
& Impact of miscommunication 
& EI for Communication, Collaboration, and Conflict Management. \\
\hline
\end{tabular}
\end{table}

\subsection{Rigour and Trustworthiness}
\label{sec:rigour}
We conceptualised and executed this study in alignment with Braun and Clarke's reflexive thematic analysis (RTA) framework \cite{braun2019reflecting, braun2021thematic}. To ensure methodological integrity, our quality control process followed Braun and Clarke's 15-point checklist for quality TA \cite{braun2021thematic}.

\begin{table}[ht]
\centering
\caption{15-point checklist for good reflexive TA}
\label{tab:checklist}
\begin{tabular}{p{3.5cm} p{8cm}}
\hline
\textbf{Process} & \textbf{Criteria} \\
\hline
1.	Transcription: & The interviews were transcribed verbatim and checked against the original audio.  \\
2-6. Coding and Theme Development: & We aimed to code in an inclusive, comprehensive, and detailed way. Themes were given the same attention and were checked against the codes, transcripts and notes. We paid particular attention to ensure consistency between the RQs, the framework-guided and reflexive analytical approach, and the final thematic interpretations. Themes were checked to ensure each one has a distinct central idea. \\

7-10. Analysis and interpretation in the written report: & We ensure the quality of our study through continuous reflexivity, prolonged engagement with the dataset, iterative analysis, and transparency. 
We moved continuously between coded extracts, full transcripts, developing themes, and our RQs to ensure our interpretations remained conceptually grounded while addressing the study's aims. \\

11. Overall: & We planned and allocated enough time between the study phases, particularly the data analysis in a way that allows us to reflect without rushing results. \\

12-15. Written Report: & Our analysis was guided by a contextualist and interpretivist perspective. We acknowledge that participants' experiences are shaped by their social and organisational contexts. We treat themes as creative, interpretive patterns actively generated through our (researchers') deep engagement with the text \cite{braun2019reflecting}. \\
\hline
\end{tabular}
\end{table}

To reflect on our active role in the research process, we exercised reflexivity, which we elaborated on in the next sub-section \ref{ref}. 

The analytical process was collaborative and intentionally rejected post-positivist measures such as coding consensus, fixed codebooks, or quantitative intercoder reliability scores, which clash with the interpretive philosophy of RTA \cite{braun2021thematic}. We did keep a shared list of codes, which changed as coding progressed and which is included in the replication package; it served as a record of our analysis.

Our collaborative analysis served as a vehicle for ``reflexive dialogue''. The initial coders met regularly to challenge each other's interpretations and explore alternative readings of the data. One more author acted as a critical companion, reviewing the developing themes and pushing the primary coders to deepen the conceptual abstraction of the patterns. 

While Goleman's emotional intelligence framework sensitised the team during study design, we analysed and generated the final themes, balancing framework-awareness with an openness to unexpected patterns driven by the participants' lived experiences. Themes were recursively reviewed and refined against the entire dataset to ensure they possessed internal coherence and distinct boundaries.

\subsubsection{Reflexivity}\label{ref}

We conducted this study using Braun and Clarke's RTA framework. This positions qualitative analysis as an inherently interpretive act shaped by the researchers' unique subjectivities, experiences, and theoretical positioning \cite{braun2019reflecting, braun2021thematic}. Following Braun and Clarke's guidelines, we treat our subjectivities as our primary analytical resources. We elaborate on them next.

Our research team brought diverse experiential lenses to the data. The first author has a mixed background in psychology, sociology and software engineering. Their deep interest in human factors and EI motivated the study's framing. The second and third authors possess extensive professional and academic experience in agile software testing, granting them immediate linguistic and contextual familiarity with the participants' day-to-day workplace challenges. The fourth author provided contextualisation as well as senior methodological oversight grounded in software quality research and industry experience.

Our backgrounds gave us ``insider status'' that accelerated data familiarisation. We understood the technical jargon, organisational structures, and sprint dynamics described by the participants. However, we recognised that this close familiarity risked ``blinding'' us to nuanced meanings. This could potentially lead us to project our own past workplace frustrations and industry assumptions onto participants' stories and experiences. For instance, our pre-existing familiarity with emotional labour in tech risked making us deductively map data onto rigid EI categories rather than remaining open to organic, messy participant experiences.

To counter this risk, we practised reflexivity. We (the first three authors) used our diverse professional backgrounds and experiences to challenge one another's readings. The fourth author, as our ``critical companion'', pushed to explore alternative interpretations and tied the findings more broadly to EI's potential in industrial practice. As a result, the final themes and discussion are interpretive constructions actively generated at the intersection of the participants' accounts, our professional subjectivities, and the theoretical frameworks that guided our field.

\subsection{Ethical Considerations}
We rigorously followed ethical protocols, providing each participant with a complete explanation of the purpose of the study and the way the collected data would be used, and we gave them the opportunity to ask questions. After they agreed to participate, we asked them to read carefully the consent form and then sign it. We then started the interview. 

For the analysis and presentation of the report, we anonymised the interviews, giving each participant an alias to ensure that their personal data was not shared.

No formal ethical review was sought. Under the Swedish Ethical Review Act (2003:460), review is required for research that processes sensitive personal data or involves a physical or psychological intervention; this study collected accounts of work experiences from consenting professionals and did not fall under either. Transcripts were stored password-protected in our university server. Participants were not compensated for their interviews.

%% file: Results.tex
\section{Results}
\label{sec:results}

This section describes the results and interpretations from the interviews.

We interviewed 16 software testers working in Sweden in different industries, including aviation, automotive, healthcare, IT services, administration, banking, and pharmaceuticals. Table \ref{tab:participants} shows their IDs, age and working experience in years. 

\begin{table}[ht]
\centering
\caption{Participants. Work experience in years. }
\begin{tabular}{lccp{2.4cm}p{2.0cm}p{1.7cm}}
\hline
\textbf{Participant} & \textbf{Age} & \textbf{Experience}  \\
\hline
P1  & 34 & 7.5  \\
P2  & 37 & 12   \\
P3  & 32 & 3    \\
P4  & 41 & 1    \\
P5  & 42 & 8    \\
P6  & 46 & 8    \\
P7  & 34 & 8    \\
P8  & 41 & 11.5 \\
P9  & 37 & 12   \\
P10 & 40 & 12   \\
P11 & 37 & 15   \\
P12 & 34 & 7    \\
P13 & 54 & 5    \\
P14 & 26 & 1.5  \\
P15 & 54 & 22   \\
P16 & 49 & 19   \\
\hline
\end{tabular}
\label{tab:participants}
\end{table}

\subsection{Insights from Interviews}
We have identified three themes from the open-ended questions in the interviews. Table \ref{tab:themes} presents the themes and sub-themes.

\begin{table}[ht]
\centering
\caption{Themes and sub-themes generated from the interviews with testers, with the number of codes under each (146 in total). Themes 2 and 3 were not divided into sub-themes.}
\label{tab:themes}
\begin{tabular}{p{7.2cm} p{4.5cm} }
\hline
\textbf{Theme} & \textbf{Sub-theme}  \\
\hline

\textbf{Theme 1: EI in Day-to-Day Testing Practice} & A. Self-regulation under pressure  \\
 & B. Positive drivers  \\

\\

\textbf{Theme 2: EI for Communication, Collaboration, and Conflict Management} & --  \\

\\

\textbf{Theme 3: Emotional Responses to Requirements Volatility} & --  \\

\hline
\end{tabular}
\end{table}

\subsubsection{\textbf{Theme 1: EI in Day-to-Day Testing Practice.}}

The importance of EI appeared most clearly in testers' accounts of their everyday work. Their comments stressed that EI was not an abstract quality but a practical tool that they can use for coping with stress, maintaining focus, and finding motivation. Participants described two main dimensions in their daily practice, where they had to face and navigate emotional states: on one side, the regulation of negative emotions under pressure, and on the other, the use of positive drivers such as clarity, recognition, and autonomy to sustain motivation and depth of testing. These two scenarios are where participants described balancing immediate emotional reactions against the longer-term goal of delivering reliable software.

\paragraph{\textbf{Sub-theme A: Self-regulation under pressure}}
Testers often encounter high-stakes moments where deadlines, shifting priorities, and interpersonal tensions generate strong emotional reactions. Instead of letting these reactions dictate their work, many described deliberate self-regulation strategies. These included taking breaks before executing critical steps, documenting decisions to reduce confusion later, and sequencing tasks to avoid mistakes when stress was highest.

For example, one participant described how high-pressure deadlines contribute to increased stress levels, particularly when exacerbated by specific actions within the team, such as "People changing all the files just before release.~(P15)".

The impact of deadlines affects not only personal emotional states but also the quality of testing, as some participants acknowledged, especially when adjustments in prioritisation are needed. Some participants mitigated this by updating managers when delays occurred. However, it was clear from other participants that maintaining software quality under such pressures increases the workload. 

Adding to the previous, interpersonal conflicts can add to the whole challenge when suggestions for improvements are misinterpreted by team members as personal criticisms, leading to immediate and visible disruptions. 

One reason participants continue to be motivated to ensure software quality despite these challenges is that they manage their expectations and understand the limits of their control over workplace factors. They also demonstrated sensitivity to others' emotional states, considering when to effectively approach a team member for assistance. For instance, a participant illustrated this sensitivity by explaining: 

\dialoguegpt{quote}{
\pquote{9}"I think it's not a good time to approach that specific person right now because I know they have a lot on their plate. So, I don't add more when they are already very stressed about something else."}

One more strategy to manage difficult situations, particularly when interacting with end users, was to remain objective to prevent stress-related errors. Testers could maintain a more professional and focused approach by avoiding unnecessary conflict and confrontation that could easily escalate. At the same time, they channelled their frustrations with the workflow into constructive action by advocating for better tool integration within the team. Such tools allowed earlier testing, helping to minimise last-minute pressures and lowering the risk of compromised quality.

Across these accounts, what participants had to manage at the same time was their own stress, their communication with colleagues and end users, and the quality of their work alongside their own well-being.

\paragraph{\textbf{Sub-theme B: Positive drivers}}
Besides the experiences from the previous sub-theme, participants also described conditions that make focused, thorough testing easier. Examples of these conditions were clear goals, predictable interfaces with product and development, and recognition for well-reasoned quality decisions. Other aspects that added to the positive drivers were having explicit expectations and stable priorities. Autonomy mattered too; being trusted to push back on unsafe shortcuts, to propose better sequencing, or to introduce lightweight tooling created a sense of ownership that carried into day-to-day choices.

Aligned with the previous, positive emotions such as happiness when reaching common goals, were also mentioned, as seen in the quote below:

\dialoguegpt{quote}{
\pquote{1}"So I think when we have these releases, if it goes well, so everybody made us happy [...] I also get happy maybe because of two things, because of the release going well and because everybody in the team is happy".}

The happiness in this situation was a driver of motivation and cohesion. It connects directly to the idea that clarity, successful outcomes, and collective recognition boost morale and encourage testers to invest in quality. Positive experiences did not remove constraints but increased persistence and care. 

The experiences participants commented on, and their act of sharing emotions within the team, were EI practices. They reacted to positive emotions and also used them to strengthen engagement and teamwork. Even without labelling these actions as EI practices, testers show how it helps them to improve their work and collaboration.

\cristysbox{ \ta{RQ1 answer:}Testers described regulating stress under deadline pressure through deliberate strategies, and drawing motivation from recognition, clarity and autonomy. We read both as EI in day-to-day practice: self-awareness and self-management on the one side, and the use of positive emotion to sustain effort on the other.}

\subsubsection{\textbf{Theme 2: EI for Communication, Collaboration, and Conflict Management}}

This theme focuses on how EI shapes how software testers communicate, collaborate, and manage conflict in agile environments. Effective communication is critical for aligning expectations, ensuring clarity of requirements, and negotiating scope when priorities shift. Testers shared some of their struggles when the communication does not flow as expected. For example, they mentioned that communication breakdowns can occur when part of the team considers testing unnecessary. Additionally, they mentioned how miscommunication among team members can lead to wasted resources and reduced morale, as expressed by one participant:

\dialoguegpt{quote}{
\pquote{11}"They had not informed us that testers were not needed despite having asked several times. Somehow, they did not inform us that we were not within the scope of their testing".}

These examples portray how unclear or inconsistent communication can strain teamwork and hinder progress in agile projects. Regarding collaboration specifically, practices such as structured documentation, delegation, and transparent reporting enable testers to coordinate efficiently and maintain trust within the team. 

Some participants advocated for using standard practices, such as establishing documentation, to alleviate individual misconceptions about project challenges. Yet, they also noted that poor documentation protocols could compromise consistency and increase the likelihood of defects in the final product. To address this, some participants introduced regular and structured reporting to promote transparency and stakeholder engagement, ensuring the work progressed effectively. Interestingly, others emphasised delegation, which streamlined tasks and built trust among team members, thereby supporting team growth.

At the same time, emotionally intelligent testers must also navigate conflict: reframing criticism, managing tensions with managers and stakeholders, and choosing when and how to raise concerns. One of the interviewees shared an experience that was significant in their work.

\dialoguegpt{quote}{
\pquote{16}"The first day of [my] assignment, they [the team] were having a test meeting. I was just thrown into that room, and the program manager said, "Hey, this is P16, he's your new test leader, because all of you are doing a crappy job, and you," addressing the test manager, "are not the test manager anymore. Bad start".}

This introduction of the participant to the team created a hostile work environment that all members had to navigate for the rest of the project. Conflict management emerged as an inseparable part of communication and collaboration in agile contexts: disagreements over priorities, critical feedback, or insufficient support often triggered emotional responses, but testers drew on EI to de-escalate, negotiate, or redirect energy productively. 

These participants' experiences with communication, collaboration, and conflict management show how EI helps them handle the demands of agile teamwork and cope with the emotional strain of misunderstandings, hostile interactions, or lack of support. At the same time, the situation described above also highlights the limits of EI in the role of software testers in hostile environments. Therefore, integrating EI with strategies that strengthen team interactions and foster EI development at the team level is important.

\cristysbox{ \ta{RQ2 answer:}Testers described delivering critical findings, handling miscommunication and de-escalating hostile situations in ways that kept trust in the team. We read these as social awareness and relationship management. P16's case marks the point where individual EI is no longer enough.}

\subsubsection{\textbf{Theme 3: Emotional Responses to Requirements Volatility}}

This theme captures the emotional responses testers experience when faced with requirements volatility. These situations often lead testers to adapt their methods or make crucial decisions regarding their testing approaches to ensure software quality. 

Testers repeatedly expressed frustration, anger, and disappointment. These negative emotions amplified the pressure of deadlines and shaped how participants prioritised work, often shifting attention toward "must-pass" tests and deferring exploratory or edge cases. These feelings were mainly triggered by changing requirements that forced them to redo work, unclear requirements (especially when interacting with people defining them) and releases that still contained unresolved bugs.

Some participants also expressed feeling lost on how to manage this type of scenario, as seen in the quote below:

\dialoguegpt{quote}{
\pquote{12}"You do not even know where to start, like what to start and how to test".} 

In contrast, others found adaptability was a natural response due to working in an agile environment, helping them continuously focus on user needs. Other participants took proactive approaches to managing this frustration, such as \textit{"going back to my manager and creating awareness of how we are currently working.~(P6)"}, demonstrating their use of negative emotions to initiate change within their teams. Finally, other testers used their disappointment as motivation to find creative solutions for the missing requirements. They did this by looking at similar systems and using those insights to continue testing.

In all these cases, emotional intensity was closely tied to decisions about coverage, prioritisation, and the balance between thoroughness and timeliness.

\cristysbox{ \ta{RQ3 answer:}Testers described frustration at requirements change turning into three kinds of decision: narrowing coverage to must-pass tests, raising the process problem with a manager, and finding workarounds from similar systems. We read the move from a recognised frustration to a stated response as EI in Goleman's sense of regulation; frustration on its own we treat as an emotional reaction.}

%% file: Discussions.tex
\section{Discussion}

Participants in this study portrayed how EI impacts the way they manage work demands. Previous work had explained how EI influences productivity, collaboration, and stress management across different software engineering roles \cite{ref5, rezvani2019emotional, ref3}. Most of that work studied developers. Three features of the testing role set our participants' accounts apart from it. A tester's main work product is a negative judgement about a colleague's work \cite{myers1979art}. That judgement is produced inside a window that upstream delays keep shortening and that the tester does not control. And the result is visible mainly when it fails, since defects that are caught leave no trace and defects that escape are traced back to the tester \cite{ref54, deak2016challenges}. Section~\ref{sec:tester-specificity} works through what each of our three themes adds to the developer-focused literature. Software testers focus on tasks that include an evaluative role where their responsibility is to question, critique, and sometimes delay releases. Hence, their interactions with the rest of the team are also different and demand specific interpersonal skills. This scenario exposes testers frequently to negative emotions such as frustration, undervaluation, and interpersonal conflict \cite{deak2016challenges, guveyi2020human}. Therefore, EI is an important tool for software testers to cope with stress, frame feedback diplomatically, and preserve trust.

In our interviews, participants described three main ways in which EI shaped their experiences:

\textbf{Buffering Stress:} Testers relied on self-awareness (they first recognised their emotional state) and self-management (applying deliberate regulation strategies like pausing before critical tasks) to regulate emotional responses to high-pressure deadlines, shifting priorities, and volatile requirements. This helped them to maintain quality despite adverse conditions, echoing Goleman's framework \cite{ref8} of EI as a foundation for resilience.

\textbf{Enabling Communication and Conflict Management:} EI supported testers' ability to foster empathy, anticipate colleagues' emotional states, and resolve conflict constructively, thus sustaining psychological safety in teams \cite{ref91, madampe2024supporting}. We read this as social awareness and relationship management in Goleman's terms: first reading a colleague's stress level (``I know they have a lot on their plate'', P9) and then de-escalating hostile situations like P16's introduction.

\textbf{Mediating Responses to Requirements Volatility:} Testers reshaped frustration caused by unclear or changing requirements into adaptive strategies, such as advocating for better documentation or leveraging experience from similar systems. This shows the potential of EI to transform emotional strain into organisational learning \cite{rezvani2019emotional}. We read this example as spanning all four of Goleman's domains: recognising frustration (self-awareness) $\rightarrow$channelling it constructively (self-management) $\rightarrow$understanding developers' constraints (social awareness) $\rightarrow$advocating for process change (relationship management).

Participants' insights expose the importance of promoting and practising EI beyond a complementary skill. EI becomes central to sustaining tester well-being and software quality by bridging individual resilience, team collaboration, and organisational improvement. Furthermore, it also impacts work productivity; for instance, a stressed tester may rush through their tasks to meet deadlines, missing potential bugs and errors. Awareness and emotion management can make the tester aware of this type of situation.

\subsection{What Testers Add to Existing Accounts of Emotion in Software Engineering}
\label{sec:tester-specificity}

Studies of emotion among developers describe emotions that track the developer's own progress: a build that fails, code that is unclear, an interruption \cite{ref5, ref3}. The emotion is directed at the developer's own task. Our participants described emotion work that is directed at other people, and that belongs to the job itself. Below we take each theme's central idea and compare it to the developer-focused literature. Table~\ref{tab:tester-contrast} summarises the contrast.

\textbf{Theme 1: Day-to-day
practice.} Developer studies report that positive affect raises productivity and that recognition follows a shipped feature \cite{ref3, ref5}. When testing works, nothing happens. When a defect reaches production, the tester is asked why it was not caught. \citet{deak2016challenges} report lack of influence and recognition as the top demotivators of testers, and \citet{ref54, ref55} describe how difficult it is to assess tester performance at all. Our participants named recognition, clear goals, and autonomy as the conditions that sustained their motivation (Theme 1, sub-theme B). We read this as the emotional mechanism behind: these are the things testers reach for when the ordinary channel for credit is closed to their role. For testers, \textbf{success is invisible and failure is attributed.}

\textbf{Theme 2: Communication and conflict.} Each defect report indicates that a colleague has made a mistake, and a tester sends several such reports daily to the same colleagues. Developer-focused approaches to empathy and engagement tracking treat emotion management as a means of maintaining shared goals \cite{madampe2024supporting, ref91}. For testers, this is a daily job requirement that entails its own "display rules". The rules regarding which feelings a worker may show on the job \cite{hochschild1983managed}. Our participants’ accounts of staying objective with end users, of avoiding confrontation that could escalate, and of choosing when to raise a concern are those display rules in practice.

Hochschild’s term for managing one’s own feelings as part of the work is emotional labour. We think this term names what Theme 2 describes more accurately than communication skill does. Consequently, the issue addressed by Theme 2 is more specific than communication in general: how a tester repeatedly communicates bad news while remaining a person whom colleagues trust. Cohen et al. \citet{ref15} describe the conflict between testers and developers as inherent to both roles and propose managing it at the process level. Theme 2 reveals the cost to the tester of managing this conflict on a day-to-day basis. For the testers, \textbf{the product of their work is a critique.}

\textbf{Theme 3: Requirements
volatility.} \citet{madampe2024supporting} studied the emotional response to requirements change across agile roles. The tester's position in that cycle is asymmetric. For a developer, a changed requirement means rewriting code. For a tester, the same change invalidates test cases, test data, environments, and regression baselines, and it shortens the time left to run them, because the release date rarely moves with the change. When asked how frustration at changing requirements affected what they tested first (question 6 in Table~\ref{tab:interview_guide}), participants described narrowing to must-pass tests and deferring exploratory work and edge cases. That time pressure by itself reduces the quality of test work is established experimentally \cite{mantyla2014time}. What our participants add is the emotional path to that trade-off: frustration at the change, then a choice about what to test first and what to leave. That is a link from an emotional state to a decision about which parts of the product go unexamined. We are not aware of a developer study that reports that link. Developers under pressure also skip work, tests they do not write and refactoring they defer, so whether the tester case differs in kind or only in what is skipped is a question for a design that compares the two roles. We consider this link (the connection between frustration and decisions regarding test coverage) to be the study's finding that most directly impacts software quality. It identifies how an emotional response influences which parts of the product are examined. For testers, \textbf{frustration converts into a coverage decision.}

Two further features of the role run through all three themes. The first is responsibility without authority. A tester can recommend delaying a release and rarely decides it. We spoke about this in Section~\ref{sec:agile-conjecture}. Adding to this, P11's account of a team that never told the testers they were out of scope shows a tester arguing for the existence of the role itself, something a developer does not have to do. The second is the number of interfaces. Testers in our sample spoke with users, support, product owners, and in the regulated domains (aviation, automotive, healthcare, banking, pharmaceuticals), with people for whom a missed defect has legal or safety weight. More interfaces mean more relationships to manage, and higher stakes raise the emotional cost of each judgement. 

\begin{table}[ht]
\centering
\caption{What each theme adds to developer-focused accounts of emotion in software engineering}
\label{tab:tester-contrast}
\small
\begin{tabular}{p{2.2cm} p{4.6cm} p{4.6cm}}
\hline
\textbf{Theme} & \textbf{What developer studies report} & \textbf{What the testers add (our reading)} \\
\hline
1. Day-to-day practice & Emotion tracks the developer's own progress; positive affect raises productivity \cite{ref3, ref5} & Recognition, clarity and autonomy are the named drivers, because a tester's success is invisible and only failure is attributed to them \cite{deak2016challenges, ref54} \\
\\
2. Communication and conflict & Empathy and commitment monitoring sustain shared team goals \cite{madampe2024supporting, ref91} & The work product is itself a criticism of a colleague, so emotion management is a daily job requirement with display rules \cite{hochschild1983managed} \\
\\
3. Requirements volatility & Change triggers an emotional cycle across agile roles \cite{madampe2024supporting} & Change invalidates test assets and shortens the remaining window, and the emotional response converts into a coverage decision: what is knowingly left untested \\
\hline
\end{tabular}
\end{table}

\subsection{From Personal Coping to Systemic Capability}
\begin{figure}[h!]
    \centering
    \includegraphics[width=1\linewidth]{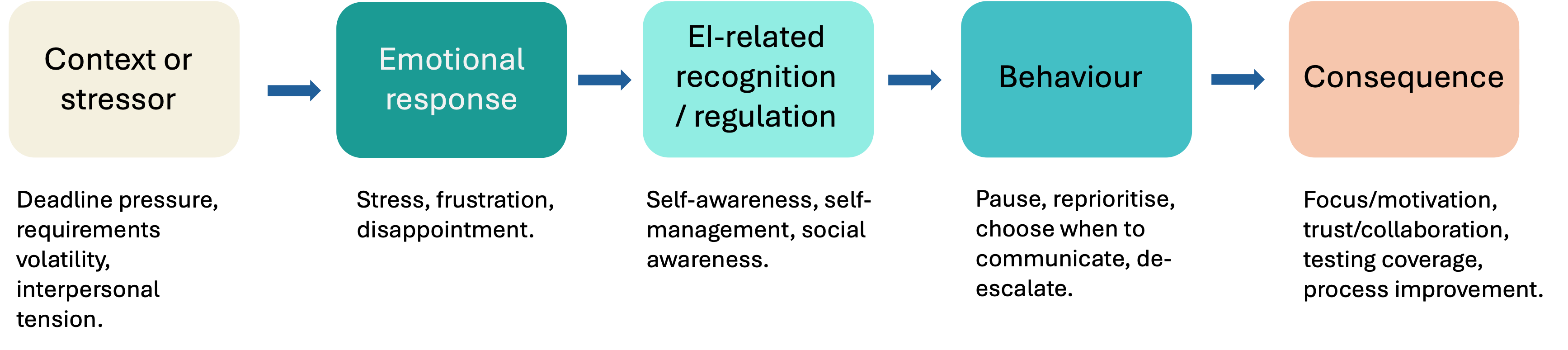}
    \caption{Interpretive sequence used to synthesise participants’ accounts of emotionally demanding situations in software testing.}
    \label{fig:minimodel}
\end{figure}

EI begins as an individual resource and skill but has the potential to scale into a collective and organisational capability. As presented in the previous subsection, at the personal level, participants described regulating emotions under pressure by pausing before executing tasks, sequencing work carefully, or documenting decisions to reduce later confusion. These micro-practices reflect Goleman's domains of self-awareness and self-management \cite{ref4}, and they can be scaled to social awareness to impact team and organisation interactions. Figure \ref{fig:minimodel} shows a common sequence in how testers practice EI and the impact on their work. An emotional response was triggered by a work-related situation; participants recognised and responded to that emotion, and the resulting behaviour influenced how they proceeded with testing. This sequence is similar to Madampe et al.'s results when handling requirements changes. They linked emotion awareness to regulation, relationship management, behavioural strategies and productivity. Our results show that a similar pattern may be present in testers' work. However, we do not have enough evidence to claim this pattern applies in general to testers beyond our sample. More work is needed with a bigger sample to propose it as a causal model.

Our results showed relational effects at the team level: empathy and timing fostered trust, while transparent documentation and structured reporting enabled coordination. At this level, testers moved beyond self-management into Goleman's domain of social awareness (actively reading colleagues' emotional states). This is illustrated by P9, who deliberately chose not to approach a stressed teammate. They also moved to relationship management by coordinating through structured reporting to maintain trust. Previous studies have linked  EI to collaboration and cohesion in agile teams \cite{ref91, madampe2024supporting}. In practice, what began as individual regulation often evolved into a shared climate of trust and predictability, enabling teams to manage miscommunication and conflict more effectively.

Regarding the organisational level, participants described channelling frustration from requirements volatility into constructive advocacy, for example, by raising awareness with managers or proposing new tools and documentation practices. This represents relationship management operating at its broadest scope in Goleman's framework, testers using emotional insight to influence organisational processes and requirements practices. Testers shifted from using EI to cope with immediate strain to enabling organisational learning, transforming personal frustration into system-level improvements. \citet{rezvani2019emotional} argued that EI fosters trust and adaptation in software projects. Our results extend this argument by showing how testers' emotional responses can reshape the requirements process itself. For example, our participants described how frustration with unclear or changing requirements motivated them to raise awareness with their managers about problematic workflows and to reconsider how testing activities were coordinated. In this sense, emotional responses became more than individual reactions to stress. These responses also acted as signals that exposed weaknesses in existing requirements and testing practices. These types of responses can potentially encourage organisational learning and process improvement.

However, EI also has challenges; for example, at the individual level, it may not be enough to overcome poorly designed processes or insufficient managerial support. While testers may individually regulate stress, de-escalate conflicts, or adapt constructively to volatile requirements, sustaining these practices requires organisational support structures. Organisational structures are crucial for embedding EI practices across teams and organisational levels. For instance,  organisations can encourage transparent communication routines, supportive management behaviours, and psychologically safe team environments. Without recognition and institutionalisation, the burden of emotional regulation falls disproportionately on testers, risking burnout and reduced effectiveness \cite{deak2016challenges, guveyi2020human}.

We advocate for organisations to consider the scaling perspective in a way that EI is fostered as an individual skill and developed as a systemic asset that organisations can institutionalise to enhance testers' well-being and software quality. In Goleman's terms, this means creating conditions where self-awareness and self-management at the individual level can grow into social awareness and relationship management at the collective level. In turn, this will make EI a team and organisational property, not just the individual testers'. For example, teams could adopt transparent communication practices, psychologically safe feedback routines, and structured ways of addressing stress, conflict, and requirements volatility. In a similar way, organisations can institutionalise EI through supportive leadership behaviours, recognition of emotional labour, and encouragement of collaborative problem-solving. This will avoid placing the burden of emotional regulation solely on individual testers. In this way, EI becomes embedded in how teams coordinate, communicate, and respond to challenges. At the same time, it contributes to more sustainable collaboration and software quality practices.

\subsection{Implications for Practice}

Below, we list implications for practitioners based on our findings and experience in the area. 

\textbf{Acknowledgement of EI as part of tester's core skills.}
Our participants described EI as what let them maintain quality under pressure. The acknowledgement of this by organisations and team leaders needs to translate into support for testers to develop the skill and to promote it within working teams. 

For example, organisations can offer training on interpersonal skills and integrate EI into feedback and evaluation processes. In addition, workshops on conflict de-escalation and empathy mapping specifically tailored to the testers. These efforts map directly onto Goleman's domains of self-awareness and self-management, which form the foundation for the interpersonal competencies that follow.

\textbf{Make hidden emotional labour visible and shareable.}
Report emotional bottlenecks in testers' work, not just defects and coverage, to make their emotional labour visible. Without recognition, this labour risks becoming an unacknowledged drain \cite{deak2016challenges}. For example, during stand-ups or team check-ins, the team leader can encourage testers to report such concerns. These meetings can be moved beyond technical blockers to include social or emotional bottlenecks. Team boards can be useful, too, for tracking recurring issues at stress points and interpersonal challenges. This practice creates opportunities for team members to recognise and acknowledge each other's emotional states, which supports the development of social awareness, part of Goleman's social competence domain.

\textbf{Institutionalise EI practices into agile routines.}
Agile teams should formally integrate emotional intelligence practices, such as structured conversations about change (for example, in requirements) and agreed-upon ways of handling conflict, so that emotional challenges are addressed at the team level, rather than being left to individuals to cope with on their own. This shifts the locus of EI from individual self-management to collective relationship management, scaling Goleman's framework beyond the person and into the team structure.

\textbf{Balance EI's protective and transformative roles.}
As a team leader or organisation, encourage testers to use EI beyond buffering stress and preventing errors, to also channel frustration into constructive improvements. Avoid relying on EI merely as a coping mechanism; instead, ensure it drives system learning, such as clearer requirements and earlier risk detection \cite{rezvani2019emotional, madampe2024supporting}. This distinction mirrors the difference between Goleman's domains, self-management (emotional self-control) and relationship management, which involves using emotional insight to drive change at the team or organisational level.

\textbf{Record what is knowingly left untested.}
When requirements changed late and the window shrank, testers in our study narrowed coverage to must-pass tests and deferred exploratory work and edge cases, and they described this as a decision made under frustration. Make that trade-off a team decision. When a change lands late, the team should state in the sprint record which tests are being dropped or deferred and why, so that what goes untested is visible to the product owner and the developers. This is the one recommendation that follows directly from Section~\ref{sec:tester-specificity}: it moves the weight of the coverage decision from the individual tester to the process.

\textbf{Train managers and developers to share the EI load.}
Managers and developers need to be trained in basic EI practices (e.g., framing feedback constructively and recognising stress cues) to distribute responsibility for trust and psychological safety \cite{ref91}. Framing feedback constructively draws on relationship management, while recognising stress cues in colleagues is a social awareness competency; both are domains that training programs can explicitly target.

\subsection{Implications for Research}
Here we propose the implications for research based on our findings.

\textbf{Test the proposed process as a model: } In our interviews, we identify a pattern where work conditions evoke emotional responses and end in behavioural answers (see Figure \ref{fig:minimodel}). More data is needed to study that process to establish whether those relationships we found occur in other settings and under what conditions. 

\textbf{Examine the relationship between emotional responses and testing coverage.} One particularly important finding is participants' description of frustration under requirements volatility becoming connected to decisions about what to test first and what to leave untested. Future research could investigate this relationship experimentally or longitudinally by examining whether emotional states and emotion-regulation strategies are associated with measurable changes in test prioritisation, exploratory testing, defect detection, or coverage. This would provide evidence on whether the relationship reported by participants translates into observable software-quality consequences.

\textbf{Measure and observe EI:} In this study, we did not measure EI; our results came from the participants' perception of their EI. Hence, future research could combine qualitative data with quantitative measurement of EI, behavioural observations, or repeated assessments of emotional states. Such studies could assess levels of EI and how testers respond to different work pressure, conflicts, and requirements volatility.

\textbf{Compare testing contexts:} Our population included manual and automated testing; however, we did not make any differences among the groups. Future studies could explore whether emotional demands and responses differ between these groups. Other different context can de explored as well. Such as regulated and non-regulated domains, and across organisational and national contexts. Particularly, it would be interesting to explore contexts that differ in autonomy, hierarchy, release pressure, or where the cost of speaking up has different implications.

\subsection{Limitations}
\label{sec:limitations}

We discuss the limitations of the study resulting from the design choices we made.

\textbf{Self-selection and recruitment bias:} Participants were recruited using personal invitations. The target population is described in Section~\ref{sec:methods-population}. People who agreed to participate likely had a particular interest in the topic or were more favourable toward emotional topics than those who declined. The target population characteristics (location, age and years of experience) served to obtain variation in the sample and not representativeness. We do not claim the sample mirrors the population of testers in Sweden.

\textbf{Leading questions and framework-shaped data:} The interview guide, particularly questions 6 and 8, contains potentially leading prompts about changing requirements and specific emotions. Because we followed previous literature to develop our guide, some topics might have led participants' answers; therefore, the findings may not be entirely participant-generated.

\textbf{EI was not measured:} In this study, we did not measure EI with a valid instrument. All the results are based on participants' perceptions of their own EI. Hence, we cannot claim or establish EI levels, nor their influence or causality in participants' behaviour or in improving software quality. Attached to this, the data are based on retrospective self-report. Therefore, results are based on how participants make sense of and remember their experiences.

\textbf{The findings are limited to the Swedish context:} The Swedish context shaped the results and participants' experiences in their work. All sixteen participants work in Sweden. Sweden has a work-life context with comparatively high autonomy and attention to work quality \cite{ref76}, and this probably shapes which positive drivers our participants named, autonomy and recognition among them, and how freely they described raising process problems with managers. We make no claim that the themes hold in settings where a tester's standing in the team is weaker or where speaking up carries a higher cost. The industries represented (aviation, automotive, healthcare, IT services, administration, banking and pharmaceuticals) give variation across regulated and unregulated domains, and readers should judge transfer to their own setting from the industries listed above and the range of experience in Table~\ref{tab:participants}. Country scope is stated in the abstract for the same reason.

\textbf{General focus:} Participants were recruited to cover both manual testing and test automation, but the study was not designed to contrast the two, and we did not analyse the data by testing type. Whether the emotional experience of testing differs between manual and automation work remains open.

%% file: Conclusion.tex
\section{Conclusion and Future Work}

We interviewed 16 software testers in Sweden about the part emotion plays in their work and generated three themes through reflexive thematic analysis. Testers described regulating stress under deadline pressure and drawing motivation from recognition, clarity and autonomy; managing the daily delivery of critical findings so that trust in the team survives; and responding to requirements change with frustration that turned into narrowed coverage, into advocacy for process change, or into workarounds. Read against developer-focused studies of emotion in software engineering, the themes point to what the testing role adds: the work product is a criticism of a colleague's work, success is invisible while failure is attributed, and the tester's window shrinks with every upstream delay. The link from an emotional state to a decision about what is left untested is, to our knowledge, not reported for developers, and we consider it the finding with the most direct bearing on software quality; the interview guide raised the topic and participants confirmed and elaborated it.

We proceeded from a conjecture, stated in Section~\ref{sec:agile-conjecture}, that once development has no separate test phase a tester's influence over quality decisions becomes largely relational, so emotional and interpersonal competence carries more weight than it did under a phased process. This study cannot test it: it has no comparison across process models. Testing it needs a design that compares testers under a phased process with testers in iterative delivery, or that follows a team through the change.

Three lines of further work follow from the study. The coverage decision deserves its own investigation: how often testers narrow coverage under late change, what is dropped, and whether making the trade-off explicit in the team changes what is dropped. The manual and automation testers in our sample were not contrasted, and the two groups plausibly differ in their exposure to end users and to maintenance load. And the study is limited to one country; the same questions asked of testers in settings where speaking up carries a higher cost would show which of the three themes travel.

%% file: references.bib
@misc{ref2,
  author = {Florea, R. and Stray, V.},
  title  = {A Global View on the Hard Skills and Testing Tools in Software Testing},
  year   = {2019},
  date   = {2019-05-25},
  note   = {Unpublished/online resource}
}

@inproceedings{ref3,
  author    = {Wr{\'o}bel, M.},
  title     = {Towards the participant observation of emotions in software development teams},
  booktitle = {Proceedings of the 2016 Federated Conference on Computer Science and Information Systems (FedCSIS)},
  year      = {2016},
  pages     = {1545--1548}
}

@book{ref4,
  author    = {Cherniss, C. and Goleman, D.},
  title     = {The Emotionally Intelligent Workplace: How to Select for, Measure, and Improve Emotional Intelligence in Individuals, Groups, and Organizations},
  address   = {San Francisco},
  publisher = {Jossey-Bass},
  year      = {2001}
}

@article{ref5,
  author  = {Girardi, D. and Lanubile, F. and Novielli, N. and Serebrenik, A.},
  title   = {Emotions and perceived productivity of software developers at the workplace},
  journal = {IEEE Transactions on Software Engineering},
  year    = {2022},
  volume  = {48},
  number  = {9},
  pages   = {3326--3341},
  doi     = {10.1109/TSE.2021.3087906}
}

@article{tracy2010qualitative,
  title={Qualitative quality: Eight “big-tent” criteria for excellent qualitative research},
  author={Tracy, Sarah J},
  journal={Qualitative inquiry},
  volume={16},
  number={10},
  pages={837--851},
  year={2010},
  publisher={Sage Publications Sage CA: Los Angeles, CA}
}

@book{braun2021thematic,
  title={Thematic Analysis: A Practical Guide},
  author={Braun, Virginia and Clarke, Victoria},
  year={2021},
  publisher={SAGE Publications},
  address={London},
  isbn={9781526417299}
}

@article{braun2019reflecting,
  title={Reflecting on reflexive thematic analysis},
  author={Braun, Virginia and Clarke, Victoria},
  journal={Qualitative research in sport, exercise and health},
  volume={11},
  number={4},
  pages={589--597},
  year={2019},
  publisher={Taylor \& Francis}
}

@article{coronado2023emotional,
  title={Emotional intelligence, leadership, and work teams: A hybrid literature review},
  author={Coronado-Maldonado, Isabel and Ben{\'\i}tez-M{\'a}rquez, Mar{\'\i}a-Dolores},
  journal={Heliyon},
  volume={9},
  number={10},
  year={2023},
  publisher={Elsevier}
}

@incollection{jordan2021managing,
  title={Managing emotions during team problem solving: Emotional intelligence and conflict resolution},
  author={Jordan, Peter J and Troth, Ashlea C},
  booktitle={Emotion and performance},
  pages={195--218},
  year={2021},
  publisher={CRC Press}
}

@incollection{ref7,
  author    = {Dahlstedt, {\AA}},
  title     = {Challenges in System Testing --- An Interview Study},
  booktitle = {Advances in Information Systems Development},
  editor    = {Nilsson, A. G. and Gustas, R. and Wojtkowski, W. and Wojtkowski, W. G. and Wrycza, S. and {\v{Z}}upan{\v{c}}i{\v{c}}, J.},
  address   = {Boston, MA},
  publisher = {Springer},
  year      = {2006},
  doi       = {10.1007/978-0-387-36402-5_89}
}

@book{ref8,
  author    = {Goleman, Daniel},
  title     = {Working with Emotional Intelligence},
  address   = {New York},
  publisher = {Bantam Books},
  year      = {1998}
}

@article{ref13,
  author  = {Salovey, P. and Mayer, J. D.},
  title   = {Emotional intelligence},
  journal = {Imagination, Cognition and Personality},
  year    = {1990},
  volume  = {9},
  pages   = {185--211}
}

@article{ref15,
  author  = {Cohen, C. F. and Birkin, S. J. and Garfield, M. J. and Webb, H. W.},
  title   = {Managing conflict in software testing},
  journal = {Communications of the ACM},
  year    = {2004},
  volume  = {47},
  number  = {1},
  pages   = {76--81}
}

@article{ref16,
  author  = {Bjarnason, E. and Borg, M.},
  title   = {Aligning Requirements and Testing: Working Together toward the Same Goal},
  journal = {IEEE Software},
  year    = {2017},
  volume  = {34},
  number  = {1},
  pages   = {20--23},
  doi     = {10.1109/MS.2017.14}
}

@misc{ref19,
  author = {Stray, V. and Florea, R. and Paruch, L.},
  title  = {Exploring human factors of the agile software tester},
  year   = {2021},
  note   = {Preprint/working paper}
}

@phdthesis{ref20,
  author = {Bar-On, Reuven},
  title  = {The development of an operational concept of psychological well-being},
  school = {Rhodes University},
  address= {South Africa},
  year   = {1988},
  note   = {Unpublished doctoral dissertation}
}

@incollection{ref23,
  author    = {Martins, J. De and Fonseca, A. and Gon{\c{c}}alves, G. and Shigemura, R. and Neto, W. and Cunha, A. and Dias, L.},
  title     = {Agile Testing Quadrants on Problem-Based Learning Involving Agile Development, Big Data, and Cloud Computing},
  year      = {2018},
  booktitle = {Lecture Notes in Computer Science},
  publisher = {Springer},
  doi       = {10.1007/978-3-319-54978-1_56}
}

@article{ref26,
  author  = {Madampe, Kashumi and Hoda, Rashina and Grundy, John},
  title   = {The emotional roller coaster of responding to requirements changes in software engineering},
  journal = {IEEE Transactions on Software Engineering},
  year    = {2022},
  volume  = {49},
  number  = {3},
  pages   = {1171--1187}
}

@article{ref27,
  author  = {Goleman, D. and Boyatzis, R. and McKee, A.},
  title   = {The emotional reality of teams},
  journal = {Journal of Organizational Excellence},
  year    = {2002},
  volume  = {21},
  pages   = {55--65},
  doi     = {10.1002/NPR.10020}
}

@article{ref28,
  author  = {Saiedian, H. and Dale, R.},
  title   = {Requirements engineering: Making the connection between the software developer and customer},
  journal = {Information and Software Technology},
  year    = {2000},
  volume  = {42},
  number  = {6},
  pages   = {419--428},
  doi     = {10.1016/S0950-5849(99)00101-9}
}

@inproceedings{ref36,
  author = {Deak, Anca.},
  title  = {A comparative study of testers motivation in traditional and agile software development},
  booktitle = {International Conference on Product-Focused Software Process Improvement},
  pages = {1--16},
  year = {2014},
  organization={Springer}
}

@article{deak2016challenges,
  title={Challenges and strategies for motivating software testing personnel},
  author={Deak, Anca and St{\aa}lhane, Tor and Sindre, Guttorm},
  journal={Information and Software Technology},
  volume={73},
  pages={1--15},
  year={2016},
  doi={10.1016/j.infsof.2016.01.002}
}

@inproceedings{ref37,
  author    = {S{\'a}nchez-Gord{\'o}n, M. and Rijal, L. and Colomo-Palacios, R.},
  title     = {Beyond Technical Skills in Software Testing: Automated versus Manual Testing},
  booktitle = {Proceedings of the IEEE/ACM 42nd International Conference on Software Engineering Workshops (ICSE-W)},
  year      = {2020},
  pages     = {161--164}
}

@book{ref48,
  editor    = {Salovey, P. and Brackett, M. A. and Mayer, J. D.},
  title     = {Emotional Intelligence: Key Readings on the Mayer and Salovey Model},
  publisher = {Dude Publishing},
  year      = {2004}
}

@inproceedings{ref54,
  author    = {Kanij, T. and Merkel, R. and Grundy, J.},
  title     = {Performance assessment metrics for software testers},
  booktitle = {Proceedings of the 5th International Workshop on Co-operative and Human Aspects of Software Engineering (CHASE)},
  address   = {Zurich, Switzerland},
  year      = {2012},
  pages     = {63--65},
  doi       = {10.1109/CHASE.2012.6223025}
}

@inproceedings{ref55,
  author    = {Kanij, T. and Merkel, R. and Grundy, J.},
  title     = {A Preliminary Survey of Factors Affecting Software Testers},
  booktitle = {Proceedings of the 23rd Australian Software Engineering Conference (ASWEC)},
  address   = {Milsons Point, NSW, Australia},
  year      = {2014},
  pages     = {180--189},
  doi       = {10.1109/ASWEC.2014.32}
}

@inproceedings{ref56,
  author    = {Kanij, T. and Merkel, R. and Grundy, J.},
  title     = {An empirical investigation of personality traits of software testers},
  booktitle = {Proceedings of the IEEE/ACM 8th International Workshop on Cooperative and Human Aspects of Software Engineering (CHASE)},
  year      = {2015},
  pages     = {1--7},
  doi       = {10.1109/CHASE.2015.7}
}

@article{ref60,
  author  = {Yardley, L.},
  title   = {Dilemmas in qualitative health psychology},
  journal = {Psychology and Health},
  year    = {2000},
  volume  = {15},
  pages   = {215--228}
}

@article{ref66,
  author  = {Priya, K. and Chandrasekar, K.},
  title   = {Relationship between Emotional Intelligence and Performance among Software Professionals in Kerala},
  journal = {International Journal of Geographical Information Science},
  year    = {2019},
  volume  = {14},
  pages   = {308--320},
  doi     = {10.26643/gis.v14i6.11841}
}

@inproceedings{ref73,
  author    = {Khan, R. and Srivastava, A. K. and Pandey, D.},
  title     = {Agile approach for Software Testing process},
  booktitle = {2016 International Conference on System Modeling \& Advancement in Research Trends (SMART)},
  address   = {Moradabad, India},
  year      = {2016},
  pages     = {3--6},
  doi       = {10.1109/SYSMART.2016.7894479}
}

@article{ref75,
  author  = {Linden, D. and Pekaar, K. and Bakker, A. and Schermer, J. and Vernon, P. and Dunkel, C. and Petrides, K.},
  title   = {Overlap Between the General Factor of Personality and Emotional Intelligence: A Meta-Analysis},
  journal = {Psychological Bulletin},
  year    = {2017},
  volume  = {143},
  pages   = {36--52},
  doi     = {10.1037/bul0000078}
}

@article{ref76,
  author  = {Gallie, D.},
  title   = {The Quality of Working Life: Is Scandinavia Different?},
  journal = {European Sociological Review},
  year    = {2003},
  volume  = {19},
  pages   = {61--79},
  doi     = {10.1093/esr/19.1.61}
}

@inproceedings{ref78,
  author    = {Hartmann, H.},
  title     = {Testers learning requirements},
  booktitle = {2014 IEEE 1st International Workshop on Requirements Engineering and Testing (RET)},
  year      = {2014},
  pages     = {12--15},
  doi       = {10.1109/RET.2014.6908672}
}

@article{ref85,
  author  = {Sayrs, L.},
  title   = {InterViews: An Introduction to Qualitative Research Interviewing},
  journal = {American Journal of Evaluation},
  year    = {1998}
}

@article{ref91,
  author  = {Gujral, H. K. and Ahuja, J.},
  title   = {Impact of Emotional Intelligence on Teamwork. A Comparative Study of Self Managed and Cross Functional Teams},
  journal = {ZENITH International Journal of Multidisciplinary Research},
  year    = {2011},
  volume  = {1},
  number  = {6},
  pages   = {1}
}

@misc{ref92,
  author = {Radha, B. and Shree, B.},
  title  = {Impact of Emotional Intelligence on Performance of Employees and Organizational Commitment in Software Industry},
  year   = {2017},
  note   = {Vol. 6, Issue 2, pp. 17--28}
}

@incollection{ref99,
  author    = {Baumgartner, M. and Klonk, M. and Mastnak, C. and Pichler, H. and Seidl, R. and Tanczos, S.},
  title     = {Role of Testers in Agile Projects},
  booktitle = {Agile Testing},
  address   = {Cham, Switzerland},
  publisher = {Springer},
  year      = {2021},
  pages     = {69--85},
  doi       = {10.1007/978-3-030-73209-7_4}
}

@article{rezvani2019emotional,
  title={Emotional intelligence: The key to mitigating stress and fostering trust among software developers working on information system projects},
  author={Rezvani, Azadeh and Khosravi, Pouria},
  journal={International Journal of Information Management},
  volume={48},
  pages={139--150},
  year={2019},
  doi={10.1016/j.ijinfomgt.2019.02.007}
}

@article{madampe2024supporting,
  title={Supporting emotional intelligence, productivity and team goals while handling software requirements changes},
  author={Madampe, Kashumi and Hoda, Rashina and Grundy, John},
  journal={ACM Transactions on Software Engineering and Methodology},
  volume={33},
  number={6},
  pages={1--38},
  year={2024},
  publisher={ACM New York, NY}
}

@inproceedings{guveyi2020human,
  title={Human factor on software quality: A systematic literature review},
  author={Guveyi, Elcin and Aktas, Mehmet S and Kalipsiz, Oya},
  booktitle={International Conference on Computational Science and Its Applications},
  pages={918--930},
  year={2020},
  organization={Springer}
}

@article{robinson2014sampling,
  author  = {Robinson, Oliver C.},
  title   = {Sampling in Interview-Based Qualitative Research: A Theoretical and Practical Guide},
  journal = {Qualitative Research in Psychology},
  year    = {2014},
  volume  = {11},
  number  = {1},
  pages   = {25--41},
  doi     = {10.1080/14780887.2013.801543}
}

@article{bar2006bar,
  title={The Bar-On model of emotional-social intelligence (ESI)},
  author={Bar-On, Reuven},
  journal={Psicothema},
  volume={18},
  pages={13--25},
  year={2006},
  publisher={Colegio Oficial de Psic{\'o}logos del Principado de Asturias}
}

@book{hochschild1983managed,
  author    = {Hochschild, Arlie Russell},
  title     = {The Managed Heart: Commercialization of Human Feeling},
  publisher = {University of California Press},
  address   = {Berkeley, CA},
  year      = {1983}
}

@inproceedings{mantyla2014time,
  author    = {M{\"a}ntyl{\"a}, Mika V. and Petersen, Kai and Lehtinen, Timo O. A. and Lassenius, Casper},
  title     = {Time pressure: a controlled experiment of test case development and requirements review},
  booktitle = {Proceedings of the 36th International Conference on Software Engineering (ICSE)},
  year      = {2014},
  pages     = {83--94},
  doi       = {10.1145/2568225.2568245}
}

@article{graziotin2018happens,
  author  = {Graziotin, Daniel and Fagerholm, Fabian and Wang, Xiaofeng and Abrahamsson, Pekka},
  title   = {What happens when software developers are (un)happy},
  journal = {Journal of Systems and Software},
  year    = {2018},
  volume  = {140},
  pages   = {32--47},
  doi     = {10.1016/j.jss.2018.02.041}
}

@book{myers1979art,
  author    = {Myers, Glenford J.},
  title     = {The Art of Software Testing},
  publisher = {John Wiley \& Sons},
  address   = {New York, NY},
  year      = {1979}
}
